\documentclass[fleqn,12pt]{article}

\usepackage{amsmath}
\usepackage{amsfonts}
\usepackage{amssymb}
\usepackage{graphics,graphicx}
\usepackage[english]{babel}
\usepackage{cite}
\usepackage{latexsym}
\usepackage[normalem]{ulem}
\usepackage[utf8]{inputenc}
\usepackage{textcomp,color}
\usepackage{graphicx}
\usepackage{amsxtra}

\usepackage{url, hyperref}

\title{Can a particle move zigzag in time?}

\author{	Sergey G. Rubin$^{1,2,a}$ }

\date{%
	$^1$  National Research Nuclear University MEPhI \\(Moscow Engineering Physics Institute), \\ 115409, Kashirskoe shosse 31, Moscow, Russia \\
	$^2$N.I. Lobachevsky Institute of Mathematics and Mechanics,\\
	Kazan  Federal  University, \\
	Kremlevskaya  street  18,  420008  Kazan,  Russia\\
	$^a$	sergeirubin@list.ru  }

\begin{document}
	
	\maketitle

\begin{abstract}
Amplitudes of quantum transitions containing time zigzags are considered. The discussion is carried out in the framework of the Minkowski metric and standard quantum mechanics without adding new postulates. It is shown that the wave function is singular at the instant of the time zigzag. Nevertheless, we argue that time zigzags are not suppressed at the quantum level, but their contribution to the amplitude is zero. The result is valid for a single particle and a non-interacting scalar field.
\end{abstract}
\section{Introduction}

The problem of the arrow of time origin has been of great interest for a long time \cite{Kiefer:2009tq,Carroll:2004pn,Aiello:2006gq,Donoghue:2020mdd}. The view that the fundamental rules of physics make no distinction between past and future is widely accepted. The common conclusion is that the presence of the time arrow is related to the macroscopic physics \cite{DiBiagio:2020jbd,PhysRevLett.123.020502}. At the same time, there are some arguments that the arrow of time has been incorporated into the quantum theory in the form of the arrow of causality \cite{Donoghue:2020mdd}.

One of the main points of discussion concerns the origin of the time arrow at the planckian energies. The idea that the notion of time is formed together with the Universe appearance and is closely related to its expansion \cite{Hartle:2008ng}, \cite{Ellis:2019mlk} looks reasonable. 
There are three questions that need to be answered. The first one is ``How was the arrow of time formed and supported during the Universe evolution?'' The impossibility of reversal motion of complicated systems is usually related to the entropy growth. But the statistical arguments cannot be applied to small systems like several particles.

The second question is ``are there any spacetimes in which closed spacetime geodesics exist?'' This subject is extensively discussed with a positive answer in papers \cite{Goldwirth:1993hx,Carlini:1996ay,Krasnikov:2001cj}  and many others. The coexistence of neighboring regions with opposite arrows of time is discussed in \cite{2001PhLA..280..239S}.

In this paper we consider the third question, ``why can a particle zigzag in space but not in time?''. Such processes have not yet been observed in the Minkowski metric, so the probability must be at least strongly suppressed \cite{2000physics..11036K}.

In this paper, we discuss the quantum transition amplitude of the time zigzag motion in the Minkowski space. We argue that although zigzag motion in time is unobservable indeed, it is not forbidden by the standard laws of quantum mechanics. The arguments why these two statements do not contradict each other are explained below.  It will be proved that the result does not depend on the time interval of particle motion in the opposite time direction. 

No complementary postulates are involved, although the path integral measure needs to be upgraded.

 {The structure of the paper is as follows. Section 2 is devoted to the study of the classical motion of a single particle. It is shown that the presence of time reversal points violates the energy conservation law. The quantum mechanical aspect is studied in section 3, where it is shown that the wave function at the time reversal points is singular. The imaginary ways to solve the problems are listed there. Section 4 is devoted to a thorough mathematical analysis of the situation based on the path integral approach. The conclusion is given in section 5.
}

\section{Classical motion}

\begin{figure}[!th]
	\centering
	\includegraphics[width=0.6\linewidth]{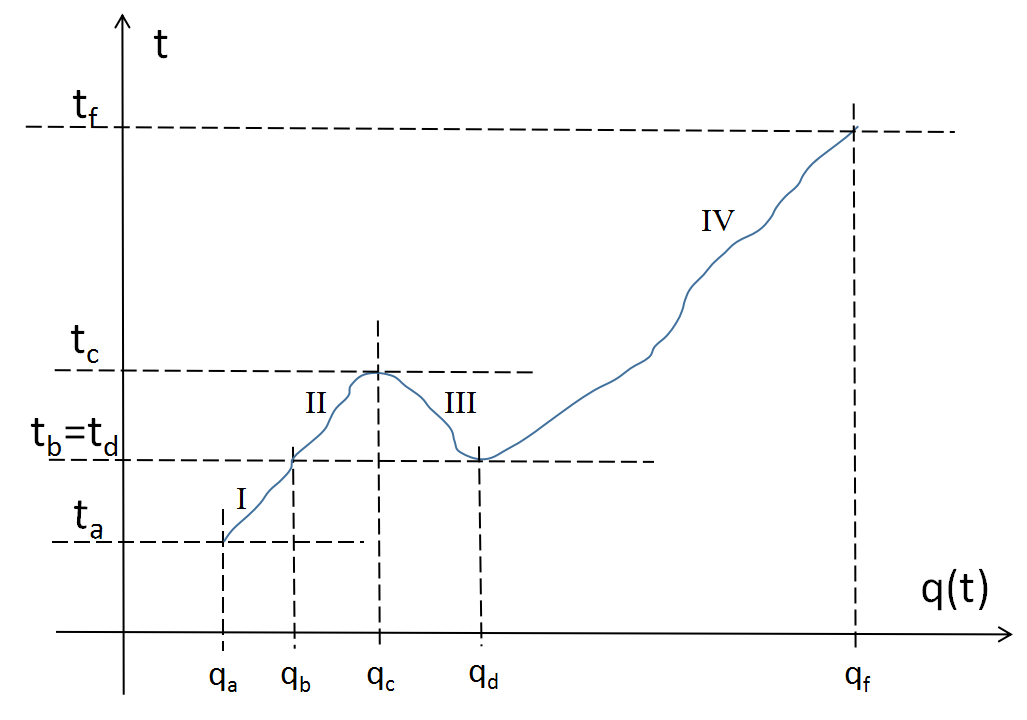}
	\caption[]{An example of trajectory. A particle moves back in time in the region III. The trajectory is parameterized by a parameter $\tau$ such that each point in the trajectory is in one-to-one correspondence with the specific value of the parameter $\tau$. Points $b$ and $d$ are characterized by different parameter values $\tau_b\neq\tau_d$ but equal time $t_b=t_d$.}
	\label{fig:traj}
\end{figure}

The situation is obvious at the classical level, but it becomes non-trivial at the quantum level. 
 To illustrate, {suppose for a moment that} the classical trajectory shown in Figure \ref{fig:traj} is a solution of the equation
	\begin{equation}
		\ddot{q}=-V'(q)
	\end{equation}
for a particle of a unit mass. {This assumption immediately leads to contradiction. Indeed,}
	the time derivative $\dot{q}$ is infinite at the moment of turning $t_c$ because $dt/dq=0$ at that point. 
Its energy
\begin{equation}
	E=\frac12 \dot{q}^2 + V(q)
\end{equation}
is also infinite and therefore a classical motion with the turning point $t_c$ is impossible. 

One can imagine a particle approaching the turning point classically, then turning back in time to a certain quantum state and moving away classically. If such a transition is possible at all, it should be suppressed, as in the case of quantum mechanical tunneling. The study presented in this paper leads to a different conclusion: a time zigzag is allowed at the quantum level, but gives zero contribution to the transition amplitude \footnote{ There is another known argument against a classical particle turning back in time. Indeed, the particle would have to cross the light cone, which is impossible for a massive particle. Nevertheless, this argument is not valid for field dynamics which is also important for this study.} .

 {The energy argument also works in the case of classical field theory. Let us identify  the scalar field $\phi (t,x)$ in an arbitrary point $x=x_P$ with the dynamical variable $q(t)$ presented in Fig. \ref{fig:traj}. The time derivative $\partial_t \phi (t,x)$ at the turning point $t=t_c$ is infinite for the same reason as for the particle motion. This means that both the equation of motion
\begin{equation}
	\partial_t^2 \phi (t,x)-\partial_x^2 \phi (t,x)=- V'(\phi)
\end{equation}	
and the energy density $\rho \propto (\partial_t \phi (t,x))^2 +...$ are singular at this moment, which forbids classical motion. Note that the time derivative $\partial_t \phi (t_c,x)$ is infinite at all points in space at the same moment $t_c$. This is true for an inertial observer in Minkowski space.}

\section{Quantum description}

 {The quantum behavior of a system is governed by the Schrödinger equation
\begin{equation}\label{Shred}
	i\hbar \frac{\partial \psi (t,\vec{x})}{\partial t}=\hat{H}\psi (t,\vec{x}).
\end{equation}
Here $\hat{H}$ is the Hamiltonian of the system.
Let us consider a possible evolution of the wave function in an arbitrary space coordinates  $\vec{x}=\vec{x}_P$. {The previous discussion can be applied in this case as well. Indeed, suppose for a moment that the solution of the equation \eqref{Shred}} $\psi (t,\vec{x}_P)$ contains turning points, as in Fig. \ref{fig:traj}, where $q(t)\equiv \psi (t,\vec{x}_P)$.}
The derivative $ {\partial_t \psi (t,\vec{x}_P)}$ tends to infinity at the turning instant $t_c$, 
\begin{equation}\label{to}
\partial_t \psi (t\to t_c,\vec{x}_P)\to\infty .
\end{equation}\label{class}
According to \eqref{Shred} the wave function is also infinite,
\begin{equation}\label{to0}
\psi (t\to t_c,\vec{x}_P)\to\infty\quad \forall \vec{x}_P
\end{equation}
 Note that the chosen point $x_P$  is arbitrary one, so that the wave function in the coordinate representation $\psi(x,t_c)$ is singular in all points $x$ at the moment $t_c$.

 {The wave function at an arbitrary time $t$ 
	\begin{equation}\label{evol}
	\psi(t>t_0)= e^{i\hat{H}(t-t_0)}\psi(t_0)
	\end{equation}
is formally expressed in terms of the evolution operator $e^{i\hat{H}(t-t_0)}$ and the wave function $\psi(t_0)$ with $t>t_0$.  In the coordinate representation, the wave functions at two points in time are connected by the kernel $K(\vec{x},t;\vec{x'},t_0)$ of the evolution operator, see e.g. the textbook \cite{Peskin:1995ev}
\begin{equation}\label{kernel}
	\psi(t,\vec{x})=\int d^3 x' K(\vec{x},t;\vec{x'},t_0) \psi(t_0,\vec{x'})
\end{equation}
}
 {Suppose that the system passes the time turning $t_c$. The next, step is to compute the wave function $\psi(t)=e^{i\hat{H}(t-t_c)}\psi(t_c)$ or in the coordinate representation
	\begin{equation}\label{kernelc}
		\psi(t,\vec{x})=\int d^3 x' K(\vec{x},t;\vec{x'},t_c) \psi(t_c,\vec{x'})
	\end{equation}
But the wave function under the integral is singular according to \eqref{to0}, so the integral in \eqref{kernelc} does not exist, and thus the wave function evolution after the turning point cannot be predicted. Also, there is no quasi-classical limit in the presence of the time reversal point.
}

 { Therefore, whether reversing time is possible or not is not clear and needs to be analyzed in more detail. One can imagine three variants of the behavior of the wave function at the turning time $t=t_c$: 
}

 {a) The energy of the system is infinite and therefore the function $\psi (t,x)$ cannot contain the time turning moments in analogy to the classical trajectories.
}

 {b) The passage of time is terminated as it happens at the singularity inside black holes or at the horizon from the point of view of a distant observer. It would mean a deep problem, which fortunately is not realized, as shown below.
}

 {c) According to the uncertainty principle, the energy at any given time cannot be defined.  Therefore, we can hope that nothing serious happened at the time $t=t_c$. In this case the turning point could be passed and it should be observed as time reversal. This seems to contradict the observations.
}

 {It can be concluded that the system behavior at the turning time $t_c$ is completely unclear and it should be analyzed in detail. For this purpose, the quantum mechanical description in the form of the path integral is used.
}

\section{Path integral approach}

\subsection{Preliminary results}

 {There are two equivalent approaches to the description of quantum systems - the Schrödinger equation and the path integral approach developed by R. Feynman \cite{feynman2010quantum}, see also \cite{kleinert2010path,Das:2019jmz}. The latter deals with transition amplitudes, 
	whose kernel has a well-known form
}
\begin{equation}\label{Kin}
	K(t_f,q_f;t_a,q_a)=\int {\cal D} q \exp\left\{i\int_{t_a}^{t_f}dt\left[\frac 12 \left(\frac{dq}{dt}\right)^2 - V(q)\right]\right\},\quad t_a < t_f,
\end{equation}
for the quantum mechanical tasks.  {Since the derivative at the turning point(s) $\dot{q}=\infty$, one might expect the same problem discussed above. But a thorough analysis below shows that integration over neighboring trajectories eliminates the singularity.
}

The summed trajectories are defined by their values $q(t_i)$ at times $t_i$, which implicitly means that $q(t)$ is a single-valued function. 
Trajectories we are interested in, do not satisfy this criterion, as it is evident from Fig. \ref{fig:traj}. Indeed, for any $t_i$ such that $t_b < t_i <t_c$ there exist three values of the function $q(t)$. 

This suggests that if we allow for both forward and backward movement in time, the known measure of integration
\begin{equation}\label{measure}
{\cal D} q ={\cal N}\prod_{i}dq(t_i).
\end{equation}
is invalidated. Here ${\cal N}$ is the normalization factor.


 {This subsection adapts the integration measure for zigzag trajectories. To simplify the problem, we will limit ourselves to the case of a single zigzag trajectory. 
}
To be more precise, when we consider such a transition amplitude, we sum only those trajectories describing a particle that starts its motion somewhere in the past, then turns backward in time at the moment $t_c$ and turns forward again at $t_d$. This means that $dt<0$ only between certain times $t_c$ and $t_d$. One of such trajectory is presented in Fig.\ref{fig:traj}. The calculation of such a transition amplitude gives a direct answer to the question posed in the title.

As a first step, we need to modernise the measure \eqref{measure}. To do this, we choose the parameter $\tau$, which varies in the interval $(\tau_a, \tau_f)$, and relate it to the time $t$ in the interval $(t_a,t_f)$. Knowledge of the turning points $\tau_c$ and $\tau_d$ greatly facilitates parameterization.
One of the way to do this is to choose a function $t(\tau)$ in the piecewise form
\begin{eqnarray}
	&t=\tau \quad &\text{at}\quad \tau<\tau_c\equiv t_c , \label{II} \\
	&t=2\tau_c-\tau  \quad &\text{at} \quad \tau_c\leq\tau\leq\tau_d \equiv 2t_c-t_d, \label{III}\\
	&t=\tau +2(\tau_c - \tau_d)
	\quad &\text{at} \quad \tau >\tau_d . \label{IV}
\end{eqnarray} 
The parameter $\tau$ grows monotonically along the trajectory so that $\tau_a <\tau_b <\tau_c <\tau_d <\tau_f .$
A particle moves backward in physical time $t$ in the interval $(\tau_c ,\tau_d)$. 
The important parameter value $\tau_b$ is defined from the condition
\begin{equation}\label{taut}
t(\tau_b)=t(\tau_d).
\end{equation}
Now we can assign a unique parameter $\tau$ to each point of the trajectory. $q(\tau)$ is a single-valued function of $\tau$ for ''one zigzag'' trajectory, so the appropriate measure in functional integral \eqref{Kin} is
\begin{equation}\label{measure2}
	{\cal D} q ={\cal N}\prod_{i}dq(\tau_i)
\end{equation}

The next step concerns the transition amplitude, which contains the turning points at instants $\tau_c$ and $\tau_d$. It can be divided into four parts
\begin{eqnarray}\label{Kaf}
&&K_Z(\tau_f,q_f;\tau_a,q_a)=\int dq_bdq_c dq_d
	K_{IV}(\tau_f,q_f;\tau_d,q_d)\times  \nonumber\\
&&K_{III}(\tau_d,q_d;\tau_c,q_c)	K_{II}(\tau_c,q_c;\tau_b,q_b)K_{I}(\tau_b,q_b;\tau_a,q_a)
\end{eqnarray} 
due to the principle of superposition. Here
the turning instants $\tau_c$ and $\tau_d$ are fixed and the time parametrization \eqref{II}-\eqref{IV} is taken into account. Remind that $\tau_b$ is obtained from the condition $t_b(\tau_b)=t_d(\tau_d)$, see Fig. \ref{fig:traj}. The transition amplitude $K_Z$ defined in \eqref{Kaf} contributes to the total amplitude \eqref{Kin}.

The amplitudes  $K_{I},K_{II},K_{III},K_{IV}$ do not have time zigzags, so they have the well-known form in terms of the new time/parameter $\tau$ due to the linear character of parametrization \eqref{II}-\eqref{IV}. 
The only difference in the  $K_{III}$ part of amplitude is the sign ''minus'' in the exponent
\begin{eqnarray}\label{KIII}
&&K_{III}(\tau_d,q_d;\tau_c,q_c)=\int \prod_{i}dq(\tau_i)\exp\left\{-iS_{III}\right\} \\
&&S_{III}=\int_{\tau_c}^{\tau_d}d\tau\left[\frac 12 \left(\frac{dq}{d\tau}\right)^2 - V(q)\right] \nonumber
\end{eqnarray}
which is the result of motion in the opposite time direction, $dt=-d\tau$ according to \eqref{III}. Sign ''$-i...$'' is important in the following deliberation. As was discussed in  \cite{Donoghue:2020mdd}, it relates to the arrow of causality. 

The central point of this study is the transition amplitude
\begin{eqnarray}\label{KKV}
	&&K_{II,III}(\tau_d,q_d;\tau_b,q_b)=\int dq_c K_{III}(\tau_d,q_d;\tau_c,q_c)K_{II}(\tau_c,q_c;\tau_b,q_b)
\end{eqnarray}
or in the path integral representation
\begin{eqnarray}\label{K23}
	&&K_{II,III}=\int dq_c\int Dq_{II}(\tau) Dq_{III}(\tau) e^{iS[q_{II}(\tau)]} e^{-iS[q_{III}(\tau)]}.
\end{eqnarray}
This amplitude consists of two pieces of trajectory - number II with time $t$ goes in forward direction, ''clockwise'' and number III with time $t$ going back in time, ''anticlockwise'', see Fig~\ref{fig:traj}. Note that the $d\tau >0$ for both pieces of trajectory. There is common point $q_c$ at $\tau=\tau_c$ where the trajectories II are finished and the trajectories III are started.  {The following section focuses on the analysis of a particle's passage through the turning point and its subsequent dynamics.
}

\subsection{Particle in an arbitrary potential}

Consider a particle motion in a potential $V(q)$ taking into account the II and III parts of the trajectory. The time intervals  $\tau_c-\tau_b$ and $\tau_d - \tau_c$ satisfy the conditions $\tau_c-\tau_b=\tau_d - \tau_c>0$ and can be parted in $N$ intervals in the standard manner
\begin{equation}\label{Ne}
	\tau_c-\tau_b =N\epsilon ,\quad \tau_d - \tau_c =N\epsilon, \quad \epsilon \to 0.
\end{equation} 

Object defined as
\begin{eqnarray}\label{near}
&&K_0(\tau_c,q'_1,q_1, n )\equiv \underset{\epsilon\to 0}{\lim} \int dq_c K_{III}(\tau_c +(n+1)\epsilon,q'_1;\tau_c+n\epsilon,q_c) \nonumber \\
&&\times K_{II}( \tau_c - n\epsilon,q_c; \tau_c-(n+1)\epsilon,q_1)
\end{eqnarray}
is important for the following discussion. Note that two time instants $\tau_c +n\epsilon$ and $\tau_c -n\epsilon$ refer to the same physical time $t$ because $\tau_c$ is the turning point. The object $K_0$ defined in \eqref{near} appears to be proportional to the $\delta$ function. Indeed, substituting the standard representation of the transition amplitude as in \eqref{KIII} into the definition \eqref{near} gives
\begin{eqnarray}
&&K_0(\tau_c,q'_1,q_1, n ) \\
&&= \underset{\epsilon\to 0}{\lim}\int dq_c\left(2\pi i\epsilon\right)^{-1/2} \exp\left[-\frac i2\frac{(q'_1-q_c)^2}{\epsilon}+i\epsilon V(q_c) +o(\epsilon^2)\right]\times \nonumber\\
&&\times \left(2\pi i\epsilon\right)^{-1/2}\exp\left[\frac i2\frac{(q_1-q_c)^2}{\epsilon}-i\epsilon V(q_c) +o(\epsilon^2)\right] \\
&&=\exp \left[\frac i2 ({q_1} ^2-{q'_1} ^2)\right] \underset{\epsilon\to 0}{\lim}\int dq_c\left(2\pi i\epsilon\right)^{-1} \exp \left[iq_c\frac{q'_1-q_1}{\epsilon}\right]=\delta (q_1-q'_1). \nonumber
\end{eqnarray}
We also put the particle mass $m=1$ and $ \hbar=1$ for not to overburden the calculations. In contrast to the typical sign in the third line, the second line contains the sign "minus"\ in the exponent, see \eqref{KIII}, so the integral in the last line strongly differs from the usual form, see \cite{Peskin:1995ev}, which is the key point.

Let us take a closer look at the amplitude in the vicinity of the turning point $\tau_c$.
\begin{eqnarray}\label{KKV2}
&&K_{II,III}(\tau_d,q_d;\tau_b,q_b)\equiv \int dq_c K_{III}(\tau_d,q_d;\tau_c,q_c)K_{II}(\tau_c,q_c;\tau_b,q_b) \nonumber\\
&&= \underset{\epsilon\to 0}{\lim}\int K_{III}(\tau_d,q_d;\tau_c+\epsilon,q''_c) dq''_c K_0(\tau_c,q''_c,q'_c,n=0)dq'_cK_{II}(\tau_c-\epsilon,q'_c;\tau_b,q_b)
\nonumber \\
&&= \underset{\epsilon\to 0}{\lim}\int  dq'_c  K_{III}(\tau_d,q_d;\tau_c+\epsilon,q'_c)K_{II}(\tau_c-\epsilon,q'_c;\tau_b,q_b).
\end{eqnarray}
Here the first line is the standard decomposition of the transition amplitude, the second line contains definition \eqref{near}. In the third line we use the fact that $K_0$ is equal to the $\delta$ function according to \eqref{near}. 

 {The key result is that the transition amplitude describing the motion in the time interval $(\tau - \epsilon, \tau + \epsilon)$ disappears. According to \eqref{KKV2} only the transition amplitudes in the intervals $\tau_b, \tau_c-\epsilon$ before the turning point and $\tau_c+\epsilon,\tau_d$ after the turning point contribute.
}

The procedure described above can be repeated at the second iteration with $n=1$:
\begin{eqnarray}\label{KKV3}
	&&K_{II,III}(\tau_d,q_d;\tau_b,q_b)= \underset{\epsilon\to 0}{\lim}\int  dq_1  K_{III}(\tau_d,q_d;\tau_c+\epsilon,q_1)K_{II}(\tau_c-\epsilon,q_1;\tau_b,q_b) \nonumber\\
	&&= \underset{\epsilon\to 0}{\lim}\int  dq_2 dq'_2 K_{III}(\tau_d,q_d;\tau_c+2\epsilon,q_2)K_0(\tau_c,q_2,q'_2,n=1)K_{II}(\tau_c-2\epsilon,q'_2;\tau_b,q_b) \nonumber\\
	&&= \underset{\epsilon\to 0}{\lim}\int  dq_2 K_{III}(\tau_d,q_d;\tau_c+2\epsilon,q_2)K_{II}(\tau_c-2\epsilon,q_2;\tau_b,q_b).
\end{eqnarray}
The second and third terms in the middle line are the  transition amplitudes acting in small time interval $\epsilon$. 

Repeating such procedure $N$ times one obtains
\begin{eqnarray}\label{KKVN}
&&K_{II,III}(\tau_d,q_d;\tau_b,q_b)
\nonumber\\
&&= \underset{\epsilon\to 0}{\lim}\int  dq_N  K_{III}(\tau_d,q_d;\tau_c+N\epsilon,q_N)K_{II}(\tau_c-N\epsilon,q_N;\tau_b,q_b)
\end{eqnarray}
According to \eqref{Ne}, $\tau_c-N\epsilon =\tau_b$ and $\tau_c+N\epsilon =\tau_d$. Both amplitudes under integral \eqref{KKVN} do not contain inverse time motion. Therefore, we can use their standard normalization
\begin{eqnarray}\label{dd}
&&K_{II}(\tau_c-N\epsilon,q_N;\tau_b,q_b)=K_{II}(\tau_b,q_N;\tau_b,q_b)=\delta(q_N-q_b), \nonumber\\
&&K_{III}(\tau_c+N\epsilon,q_N;\tau_d,q_d)=K_{III}(\tau_d,q_N;\tau_d,q_d)=\delta(q_N-q_d) \nonumber
\end{eqnarray}
to substitute it into \eqref{KKVN} that leads to the following result
\begin{equation}\label{Kd}
	K_{II,III}(\tau_d,q_d;\tau_b,q_b)=\delta(q_b-q_d),\quad \tau_b\neq \tau_d .
\end{equation} 
This result is valid if a time reverse motion is assumed in the interval $(\tau_b , \tau_d)$. 

Finally, we substitute \eqref{Kd} by \eqref{Kaf} to obtain the transition amplitude $\tau_a,q_a\to \tau_f,q_f$ in the form
\begin{eqnarray}\label{Kaf2}
	&&K(\tau_f,q_f;\tau_a,q_a)=i\int dq_b
	K_{IV}(\tau_f,q_f;\tau_b,q_b)K_{I}(\tau_b,q_b;\tau_a,q_a)
\end{eqnarray}
The expression \eqref{Kaf2} is the standard form for the quantum transition amplitude for a particle moving ''clockwise'' in the ordinary time regime. The part containing the reverse motion completely disappears, regardless of the potential shape and the duration of the time interval. 

The discussion above shows that the reverse motion is feasible, but utterly unobservable. This conclusion is the result of calculations based on the standard quantum mechanics.
The application of this method to Lagrangians containing higher derivatives is not so obvious and deserves further discussion.

\subsection{Free massive scalar field}
The result of the previous section can be easily applied to the scalar field with action
\begin{equation}\label{Ss}
	S=\frac 12\int d^4x \left[\partial_{\mu} \phi\partial^{\mu} \phi-m^2\phi^2\right]
\end{equation}
Suppose that a space region $\cal{V}$ of finite size contains a certain field configuration $\phi(t_{in},x),\, x\in \cal{V}$ at time $t_{in}$. Can this field turn backwards in time? 

In the momentum representation, action \eqref{Ss} describes the set $\{\phi_p\}$ of the harmonic oscillators
\begin{equation}\label{Sscal}
S=\sum_{{\bf p}}\frac 12\int dt \left[\dot{\phi}_p(t)^2 - ({\bf p}^2 +m^2)\phi_p^2(t)\right] .
\end{equation}
Each mode $\phi_p(t)$ evolves independently and hence it can be considered as a set of independent particles acting in the potential $V(\phi_p) =({\bf p}^2+m^2)\phi_p$. Therefore, we can apply the results of the previous subsections. The time zigzags of the scalar field are possible, but they are unobservable since they do not affect the transition amplitude.

\section{Conclusion}

We studied the possibility of a zigzag motion in time for a particle obeying the standard laws of quantum mechanics.
The transition amplitudes, which include the reverse (zigzag) motion in time, were calculated using the standard path integral approach.
Correct determination of the path integral measure is required to account for time zigzags. 

The primary finding is that time zigzags are not prohibited at the quantum level.  {Singularities at time reversal points are smoothed by integration over neighboring trajectories. In addition, the time interval between turning points can be arbitrarily large without leaving any traces in the transition amplitude. This result is not consistent with the assumptions made at the end of section 3.
}

This conclusion is also applicable to free scalar fields considered as a set of non-interacting oscillators. Theories with higher derivatives need further analysis.

\section{Acknowledgments}
The author is grateful to O. Zaslavskii for his interest in this research.
The work was funded by the Ministry of Science and Higher Education of the Russian Federation, Project "New Phenomena in Particle Physics and the Early Universe" FSWU-2023-0073
and the Kazan Federal University Strategic Academic Leadership Program.


\begin{thebibliography}{10}


	
	\bibitem{Kiefer:2009tq}
	Claus Kiefer.
	\newblock {Does time exist in quantum gravity?}
	\newblock {\em Einstein Stud.}, 13:287--295, 2017.
	\newblock \href {http://arxiv.org/abs/0909.3767} {\path{arXiv:0909.3767}},
	\href {https://doi.org/10.1007/978-1-4939-3210-8_10}
	{\path{doi:10.1007/978-1-4939-3210-8_10}}.
	
	\bibitem{Carroll:2004pn}
	Sean~M. Carroll and Jennifer Chen.
	\newblock {Spontaneous inflation and the origin of the arrow of time}.
	\newblock 10 2004.
	\newblock \href {http://arxiv.org/abs/hep-th/0410270}
	{\path{arXiv:hep-th/0410270}}.
	
	\bibitem{Aiello:2006gq}
	Matias Aiello, Mario Castagnino, and Olimpia Lombardi.
	\newblock {The Arrow of time: From universe time-asymmetry to local
		irreversible processes}.
	\newblock {\em Found. Phys.}, 38:257--292, 2008.
	\newblock \href {http://arxiv.org/abs/gr-qc/0608099}
	{\path{arXiv:gr-qc/0608099}}, \href
	{https://doi.org/10.1007/s10701-007-9202-0}
	{\path{doi:10.1007/s10701-007-9202-0}}.
	
	\bibitem{Donoghue:2020mdd}
	John~F. Donoghue and Gabriel Menezes.
	\newblock {Quantum causality and the arrows of time and thermodynamics}.
	\newblock {\em Prog. Part. Nucl. Phys.}, 115:103812, 2020.
	\newblock \href {http://arxiv.org/abs/2003.09047} {\path{arXiv:2003.09047}},
	\href {https://doi.org/10.1016/j.ppnp.2020.103812}
	{\path{doi:10.1016/j.ppnp.2020.103812}}.
	
	\bibitem{DiBiagio:2020jbd}
	Andrea Di~Biagio, Pietro Don\`a, and Carlo Rovelli.
	\newblock {The arrow of time in operational formulations of quantum theory}.
	\newblock {\em Quantum}, 5:520, 2021.
	\newblock \href {http://arxiv.org/abs/2010.05734} {\path{arXiv:2010.05734}},
	\href {https://doi.org/10.22331/q-2021-08-09-520}
	{\path{doi:10.22331/q-2021-08-09-520}}.
	
	\bibitem{PhysRevLett.123.020502}
	P.~M. Harrington, D.~Tan, M.~Naghiloo, and K.~W. Murch.
	\newblock Characterizing a statistical arrow of time in quantum measurement
	dynamics.
	\newblock {\em Phys. Rev. Lett.}, 123:020502, Jul 2019.
	\newblock URL: \url{https://link.aps.org/doi/10.1103/PhysRevLett.123.020502},
	\href {https://doi.org/10.1103/PhysRevLett.123.020502}
	{\path{doi:10.1103/PhysRevLett.123.020502}}.
	
	\bibitem{Hartle:2008ng}
	James~B. Hartle, S.~W. Hawking, and Thomas Hertog.
	\newblock {The Classical Universes of the No-Boundary Quantum State}.
	\newblock {\em Phys. Rev. D}, 77:123537, 2008.
	\newblock \href {http://arxiv.org/abs/0803.1663} {\path{arXiv:0803.1663}},
	\href {https://doi.org/10.1103/PhysRevD.77.123537}
	{\path{doi:10.1103/PhysRevD.77.123537}}.
	
	\bibitem{Ellis:2019mlk}
	George F.~R. Ellis and Barbara Drossel.
	\newblock {Emergence of time}.
	\newblock {\em Found. Phys.}, 50(3):161--190, 2020.
	\newblock \href {http://arxiv.org/abs/1911.04772} {\path{arXiv:1911.04772}},
	\href {https://doi.org/10.1007/s10701-020-00331-x}
	{\path{doi:10.1007/s10701-020-00331-x}}.
	
	\bibitem{Goldwirth:1993hx}
	Dalia~S. Goldwirth, Malcolm~J. Perry, Tsvi Piran, and Kip~S. Thorne.
	\newblock {The Quantum propagator for a nonrelativistic particle in the
		vicinity of a time machine}.
	\newblock {\em Phys. Rev. D}, 49:3951--3957, 1994.
	\newblock \href {http://arxiv.org/abs/gr-qc/9308009}
	{\path{arXiv:gr-qc/9308009}}, \href
	{https://doi.org/10.1103/PhysRevD.49.3951}
	{\path{doi:10.1103/PhysRevD.49.3951}}.
	
	\bibitem{Carlini:1996ay}
	A.~Carlini and I.~D. Novikov.
	\newblock {Time machines and the principle of selfconsistency as a consequence
		of the principle of stationary action. 2. The Cauchy problem for a
		selfinteracting relativistic particle}.
	\newblock {\em Int. J. Mod. Phys. D}, 5:445--480, 1996.
	\newblock \href {http://arxiv.org/abs/gr-qc/9607063}
	{\path{arXiv:gr-qc/9607063}}, \href
	{https://doi.org/10.1142/S021827189600028X}
	{\path{doi:10.1142/S021827189600028X}}.
	
	\bibitem{Krasnikov:2001cj}
	S.~Krasnikov.
	\newblock {The Time travel paradox}.
	\newblock {\em Phys. Rev. D}, 65:064013, 2002.
	\newblock \href {http://arxiv.org/abs/gr-qc/0109029}
	{\path{arXiv:gr-qc/0109029}}, \href
	{https://doi.org/10.1103/PhysRevD.65.064013}
	{\path{doi:10.1103/PhysRevD.65.064013}}.
	
	\bibitem{2000physics..11036K}
	L.~Ya. {Kobelev}.
	\newblock {Why We Can Not Walk To and Fro in Time as Do it in Space? (Why the
		Arrow of Time is Exists?)}.
	\newblock {\em arXiv e-prints}, page physics/0011036, November 2000.
	\newblock \href {http://arxiv.org/abs/physics/0011036}
	{\path{arXiv:physics/0011036}}.
	
	\bibitem{2001PhLA..280..239S}
	L.~S. {Schulman}.
	\newblock {Resolution of causal paradoxes arising from opposing thermodynamic
		arrows of time}.
	\newblock {\em Physics Letters A}, 280(5-6):239--245, March 2001.
	\newblock \href {http://arxiv.org/abs/cond-mat/0102014}
	{\path{arXiv:cond-mat/0102014}}, \href
	{https://doi.org/10.1016/S0375-9601(01)00093-7}
	{\path{doi:10.1016/S0375-9601(01)00093-7}}.
	
	\bibitem{Das:2019jmz}
	Ashok Das.
	\newblock {\em {Field Theory}: {A Path Integral Approach}}.
	\newblock WSP, 2019.
	\newblock \href {https://doi.org/10.1142/11339} {\path{doi:10.1142/11339}}.
	
	\bibitem{Peskin:1995ev}
	Michael~E. Peskin and Daniel~V. Schroeder.
	\newblock {\em {An Introduction to quantum field theory}}.
	\newblock 1995.
	\newblock URL:
	\url{http://www.slac.stanford.edu/spires/find/books/www?cl=QC174.45%3AP4}.
	
	\bibitem{feynman2010quantum}
	Feynman, R.P. and Hibbs, A.R. and Styer, D.F.
	\newblock {\em 	{Quantum Mechanics and Path Integrals}},
	\newblock {2010},
	\newblock{Dover Publications},
	\newblock URL:
	\url{https://books.google.es/books id=JkMuDAAAQBAJ},
	
	
\bibitem{kleinert2010path}
Kleinert, Hagen
\newblock {\em {Path Integrals in Quantum Mechanics, Statistics, Polymer Physics, and Financial Markets}},
\newblock {2010},
\newblock{Dover Publications},

	
\end{thebibliography}

\end{document}